\documentclass[aps,pra,twocolumn,superscriptaddress,floatfix]{revtex4-1}
\usepackage{epsfig,amsmath,amssymb,color}
\usepackage{bm}
\usepackage{tensor}
\begin{document}

\title{Fresnel-Fizeau drag: Invisibility conditions for all inertial observers}
\author{Jad C. Halimeh}
\affiliation{Physics Department and Arnold Sommerfeld Center for Theoretical Physics, Ludwig-Maximilians-Universit\"at M\"unchen, 80333 M\"unchen, Germany}
\author{Robert T. Thompson}
\affiliation{Department of Mathematics and Statistics, University of Otago, Dunedin 9054, New Zealand}

\begin{abstract}
It was recently shown [Halimeh \emph{et al.} arXiv:1510.06144 (to appear in Phys. Rev. A)] that as a result of the Doppler effect, inherently dispersive single-frequency ideal free-space invisibility cloaks in relative motion to an observer can only cloak light whose frequency in the cloak frame coincides with the operational frequency of the cloak, although an infinite number of such rays exist for any cloak motion. In this article, we show analytically and through ray-tracing simulations that even though this relationship can be relaxed by simplifying the ideal invisibility cloak into a broadband amplitude cloak, Fresnel-Fizeau drag uncloaks the phase of light in the inertial frame of the cloak thereby compromising its amplitude cloaking in all other inertial frames. In other words, only an invisibility device that perfectly cloaks both the amplitude and the phase of light in its own inertial frame will also (perfectly) cloak this light in any other inertial frame. The same conclusion lends itself to invisible objects that are not cloaks, such as the invisible sphere.
\end{abstract}

\maketitle

\section{Introduction}
A staple of fantasy and science fiction, cloaking devices allow the cloaked magician, wizard, or spaceship to move about undetected.
The ideal cloaking device would hide an object from the view of any observer when illuminated by any source, and would itself be invisible to any external observer.
We argue here that such a cloak must unfortunately remain forever within the realm of fantasy and fiction, and that the more prosaic reality of invisibility devices is fundamentally limited by special relativity.

The scientific study of cloaking devices began with the mathematical manipulations of what has become known as transformation optics or transformation electromagnetics \cite{Leonhardt2006a,Pendry2006,Leonhardt2006b,Schurig2006a}.
In this approach, Maxwell's equations are manipulated through a suitable mathematical operation (transformation) such that the path of propagating light is diverted around a region and continues on the other side as if the region were not there; as if it were invisible.
This operation simultaneously derives the required properties of the cloak.

\begin{figure}[h]
\includegraphics[scale=.4]{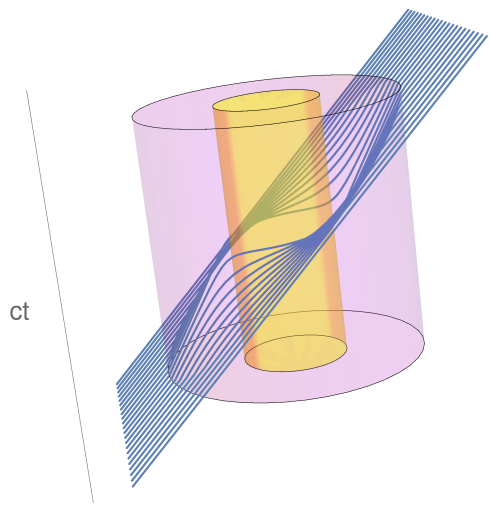}
\caption{(Color online) In the Minkowski vacuum, concentric cylinders indicate the time-like world-tube of a two-dimensional slice of a cylindrical cloak. Outside the cloak, light-like rays travel on straight lines at $45^{\circ}$.  Inside the cloak their paths vary from time-like to space-like (indicating regions of faster than light propagation) but their average behavior is null.  Throughgoing rays exit the cloak at the same space-time events as would uncloaked, unobstructed rays.}
\label{Fig:4DCloak}
\end{figure}

One quickly runs into difficulty when trying to build a cloaking device.
If the transformation is purely spatial, as in \cite{Leonhardt2006a,Pendry2006} and many subsequent papers, then the resulting mathematical description of the cloak lives up to the fictional standard; light avoids the cloak cavity and emerges on the other side at the same position and time as would light traveling through vacuum.
In such a cloak, both the amplitude and phase of light passing through the cloak are preserved relative to uncloaked light irrespective of light frequency, and we refer to this as a ``full-spectrum perfect cloak.''
But since the spatial path through the cloak is longer than the uncloaked distance, the speed of light through the cloak must be larger than the vacuum speed of light $c$, and thus in reality the full-spectrum perfect cloak cannot cloak signals without violating causality \cite{Sommerfeld1907,Craeye2012,Monticone2013,Monticone2014}.  In other words, the trajectory of light in the cloak becomes space-like, as depicted in Fig.\ \ref{Fig:4DCloak}.

There is an exception to causality violation: The indistinguishability of wavefronts in a pure single-frequency wave implies its speed is not restricted by causality, but it also cannot carry a signal.
Such single-frequency cloaks \cite{Pendry2006} are physically realizable through the use of resonant metamaterials \cite{Schurig2006b}, but by causality and the Kramers-Kronig relation such single-frequency cloaks are heavily dispersive \cite{Craeye2012,Monticone2013,Monticone2014}. We refer to them as ``single-frequency perfect cloaks'' because any such cloak perfectly cloaks both the amplitude and the phase of light carrying the right frequency -- the operational frequency of this cloak -- in its inertial frame.

One may attempt to circumvent causality violation by allowing light to take extra time to get through the cloak, which is equivalent to including a time transformation in addition to a spatial transformation \cite{Leonhardt2009,Perczel2011,Tyc2010,Xu2013,Chen2013,Thompson2015}.  
In this case a light ray exits the cloak at the same spatial position as would light traveling through vacuum, but at a later time relative to an uncloaked ray.
Such a cloak preserves the amplitude of throughgoing light but does not preserve the phase, and we refer to such cloaks as ``amplitude cloaks.''
The trajectories of light through an amplitude cloak are time-like, so causality is respected and the cloak may have broadband operationality.
Since the phase is irrelevant for incoherent natural light, amplitude cloaking is an attractive solution.

Thus there are two reasonable, if not mathematically ideal, options for physically realizable cloaks: single-frequency perfect cloaks and amplitude cloaks.
The next question is whether either type of cloak functions as desired by allowing a cloaked wizard to move about undetected.
It turns out that both categories of cloaks are betrayed by simple physical effects of special relativity in the presence of relative motion between cloak and source or detector.

For single-frequency perfect cloaks, detailed investigations have recently shown that the Doppler shift between source and moving cloak may expose the cloak's presence \cite{Halimeh2016a}.
Simply put, the single-frequency perfect cloak will not work if the Doppler-shifted frequency is not equal to the operational frequency, and the amount of scattering increases in proportion to the relative velocity between reference source and cloak.
However, one may fine tune the source frequency, cloak speed, and ray direction such that the Doppler-shifted frequency is equal to the operational frequency of the cloak in the cloak frame \cite{Halimeh2016a}. Furthermore, the cloak becomes fundamentally non-reciprocal in the sense that light entering the cloak may be red(blue)-shifted to the cloak's operational frequency and pass through as desired, but if that light is subsequently retroreflected and passes through the cloak in the opposite direction it will be blue(red)-shifted even further away from the operational frequency.

We show here that Fresnel-Fizeau drag \cite{Fresnel1818,Fizeau1851,Jackson1999} of light by amplitude cloaks in relative motion causes fundamental image distortions.
While a single-frequency perfect cloak in motion can be fine tuned by adjusting the operational or source frequency, fine tuning an amplitude cloak to compensate for Fresnel-Fizeau drag would, if at all possible, require a complete redesign of the cloak to such an extent that it would no longer function as desired in the rest frame of the cloak, nor in any other frame.
Image distortions induced by an amplitude cloak in relative motion are therefore unavoidable in all but a single frame of reference.
This same conclusion immediately lends itself to non-cloak invisibility devices, such as the invisible sphere \cite{Minano2006}, and to other broadband transformation-optics-based devices.

In Sec.\ \ref{Sec:FresnelDrag} we show that Fresnel-Fizeau drag in an amplitude cloak introduces image distortions, except for the special case of light propagation parallel to cloak velocity, for all but a single choice of inertial observer. In Sec.\ \ref{Sec:Simulations} we illustrate the extent of such distortions for a range of  velocities by ray tracing both the cylindrical cloak \cite{Pendry2006,Schurig2006a} and the invisible sphere \cite{Minano2006}.
We conclude in Sec.\ \ref{Sec:Conclusions}.

\begin{figure}
	\includegraphics[scale=.9]{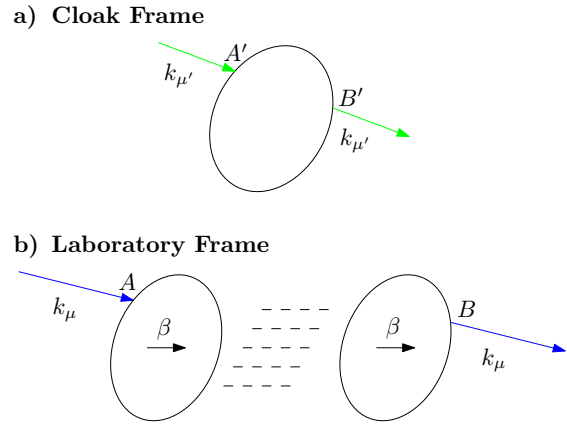}
	\caption{(Color online) a) In the cloak frame, outgoing rays suffer a time delay but no spatial shift relative to an uncloaked ray. b) A ray in the laboratory frame is Fresnel-Fizeau dragged by the moving cloak.  Lorentz transforming from cloak to laboratory, the time interval in the cloak frame is mixed into a combination of time and space intervals. For this specific illustration, the light ray is Doppler red-shifted going from the laboratory frame to the cloak frame as the cloak velocity $\vec{\beta}$ and the the light spatial wave vector $\vec{k}$ share the same direction along the $x$-axis in the laboratory frame. However, the same deterioration in cloaking would occur even if the cloak were traveling in the opposite direction. }
	\label{Fig:MovingCloak}
\end{figure}

\section{Fresnel-Fizeau drag in cloaks} \label{Sec:FresnelDrag}
To illustrate the Fresnel-Fizeau drag \cite{Fresnel1818,Fizeau1851,Jackson1999} of light passing through a cloaking device, we consider a cloak moving with speed $\beta$ in the positive $x$-direction relative to the coordinate frame of laboratory observer $S$.
Initialize the system in frame $S$, where light rays are emitted from the points $x^{\mu}_0$ and propagate with wave vector $k_{\mu}$ until they are detected at some later event on the imaging or detection plane $y=y_d$.  
We therefore imagine shining a light source at the surface $y=y_d$ and ask how the image on the plane is shifted or distorted as a cloak passes by, as in Fig.\ \ref{Fig:MovingCloak}.

These rays are to be understood as the geometric optics limit description of the path traversed by a wavefront.
In broadband systems, such as vacuum or inside the amplitude cloak, we may also think of the ray as the trajectory of a spatially confined multifrequency pulse of light, such as a laser pulse.

Our analysis makes no assumption on the direction of light propagation nor on the details of the cloak geometry or construction, but we assume that the combination of wave four-vector and cloak motion satisfies the conditions elucidated in Ref. \cite{Halimeh2016a} to ensure that the frequency in the cloak frame falls within the cloak's operational bandwidth.
In particular, if $k_{\mu}=(-\omega/c,\vec{k})$ and $\beta^{\mu} = c\gamma(1,\vec{\beta})$ are the wave four-vector of incident light and four-velocity of the cloak, relative to the laboratory frame, then the frequency $\omega'$ measured in the cloak frame must lie within the operational bandwidth of the cloak $\Delta\omega'_o=[\omega'_{\text{min}},\omega'_{\text{max}}]$
\begin{equation} \label{Eq:DispersionCondition}
 \omega' = -k_{\mu}\beta^{\mu} \in \Delta\omega'_o.
\end{equation}
For a given source frequency and operational bandwidth one may always find wave vectors such that this is true, by satisfying the condition
\begin{equation}
\frac{1}{c\gamma}(\gamma\omega -\omega'_{\text{max}}) < \vec{k}\cdot\vec{\beta} < \frac{1}{c\gamma} (\gamma\omega - \omega'_{\text{min}}).
\end{equation}
In the ray tracing simulations of Sec.\ \ref{Sec:Simulations} we ensure that $\omega$ and $\vec{k}$ satisfy this condition for each cloak velocity under investigation.
\subsection{Light propagation through the vacuum}
Let light propagate with wave four-vector $k_{\mu}$.  In vacuum, the tangent four-vector \cite{Misner1973} of the associated ray is
\begin{equation} \label{Eq:VacuumTangent}
 v^{\mu} = \frac{dx^{\mu}}{d\tau} = \eta^{\mu\nu}k_{\nu},
\end{equation}
where $\tau$ measures the affine time of the ray rather than the coordinate time of an observer (as defined here, the affine time is a dimensionless parametrization of the curve), $\eta^{\mu\nu}$
is the Minkowski metric of flat spacetime, and the condition
\begin{equation}
  v^{\mu}k_{\mu} = 0
\end{equation} 
gives a relationship between the frequency $\omega$ and spatial wave vector $\vec{k}$.
The location of the wavefront after a duration of affine time $\Delta\tau$ is subsequently found by integrating the tangent four-vector
\begin{equation}
\begin{aligned}
x^{\mu}(\tau) & = \int v^{\mu} d\tau + x^{\mu}_0 \\
              & = v^{\mu}\Delta\tau + x^{\mu}_0.
\end{aligned} 
\end{equation}

Consider now an observer $S'$ moving with speed $\beta$ in the positive $x$-direction of the laboratory frame, and let the origins of $S$ and $S'$ momentarily coincide.
Lorentz transforming the tangent vector from $S$ to $S'$ we find the position of the pulse after affine time $\Delta\tau$ as measured by $S'$ is related to the position measured by $S$ through the standard Lorentz transformation $\Lambda\indices{^{\mu'}_{\mu}}$ as expected \cite{Misner1973}:
\begin{equation}
 \begin{aligned} 
 x^{\mu'} & = \int \Lambda\indices{^{\mu'}_{\mu}} v^{\mu} d\tau + \Lambda\indices{^{\mu'}_{\mu}}x^{\mu}_0 \\
          & = \Lambda\indices{^{\mu'}_{\mu}} v^{\mu} \Delta\tau + \Lambda\indices{^{\mu'}_{\mu}}x^{\mu}_0 \\
          & = \Lambda\indices{^{\mu'}_{\mu}} x^{\mu}.
 \end{aligned} 
\end{equation}

\begin{figure}
	\includegraphics[scale=.9]{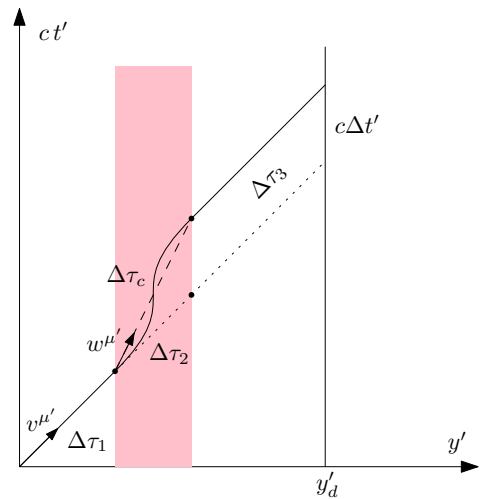}
	\caption{(Color online) In the cloak frame, a light ray (solid curve) traverses the world tube (pink-shaded region) of an amplitude cloak such that the ray tangent is always time-like in the cloak.  The detailed behavior of the ray through the cloak may be replaced by the average ray behavior with tangent vector $w^{\mu'}$ (dashed line). The image of the cloaked ray on imaging plane $y'_d$ is shifted in time relative to the uncloaked ray (dotted line).  This time delay manifests as a combined time delay and spatial shift in any other inertial frame.}
	\label{Fig:1+1Cloak}
\end{figure}

Keeping in mind the goal of analyzing the outcome when a cloaking device moves through the scene, let us break up the duration of affine time into three parts $\Delta\tau = \Delta\tau_1 + \Delta\tau_2 + \Delta\tau_3$ corresponding to the duration of affine time $\Delta\tau_1$ required for the pulse to get from the emission point to the point where it would enter the cloak; the affine time $\Delta\tau_2$ required for the pulse to traverse the region that will become occupied by the cloak; and the affine time $\Delta\tau_3$ required for the pulse to traverse the final distance from where it would exit the cloak to the imaging plane, as depicted in Fig.\ \ref{Fig:1+1Cloak}.

\subsection{Passing cloak}
Now consider a cloaking device passing through the laboratory setup. 
The defining characteristic of a cloaking device is that the outgoing ray continues in the same spatial direction as the ingoing ray, so the outgoing tangent four-vector is identical to the ingoing tangent four-vector.
Thus by the fundamental theorem of calculus we may ignore the detailed behavior of the integrated tangent four-vector through the cloak and replace it with the integrated behavior of an average tangent four-vector $w^{\mu}$, thereby connecting the entry and exit events with a straight line representing the average behavior of the ray inside the cloak, as in Fig.\ \ref{Fig:1+1Cloak}.
Relative to the entry point, the spacetime event of the exit point is $w^{\mu}\Delta\tau_c$, where $\Delta\tau_c$ is the amount of affine time required for the light to traverse the cloak.

Integrating the tangent four-vector in the cloak frame from the point of emission, through the cloak, to the imaging plane, the new imaging event is
\begin{equation} \label{Eq:ComovingDetection}
\begin{aligned} 
 \bar{x}^{\mu'} & = v^{\mu'}\Delta\tau_1 + w^{\mu'}\Delta\tau_c + v^{\mu'}\Delta\tau_3 + x^{\mu'}_0 \\
          & = \Delta\tau v^{\mu'} + x^{\mu'}_0 + w^{\mu'}\Delta\tau_c - v^{\mu'}\Delta\tau_2 \\
          & = x^{\mu'} + w^{\mu'}\Delta\tau_c - v^{\mu'}\Delta\tau_2 \\
          & = x^{\mu'} + \Delta x^{\mu'}.
 \end{aligned}
\end{equation}
The imaging event as described in the laboratory frame is obtained with the Lorentz transformation
\begin{equation} \label{Eq:LabDetection}
  \begin{aligned}
   \bar{x}^{\mu} & = \Lambda\indices{^{\mu}_{\mu'}}\bar{x}^{\mu'} \\
                 & = x^{\mu} + \Lambda\indices{^{\mu}_{\mu'}}\Delta x^{\mu'} \\
                 & = x^{\mu} + \Delta x^{\mu}.
  \end{aligned}
\end{equation}

Thus the imaging event of the cloaked ray may differ from the imaging event of the uncloaked ray, and if it differs in one frame it will differ in all frames.
If the cloak behaves as an amplitude cloak in its rest frame then $\Delta x^{\mu'}$ is simply a time delay, but since the Lorentz transformation mixes time intervals with space intervals, the laboratory observer will measure a spatial shift unless $\Delta x^{\mu'}=0$.
From Eq.\ (\ref{Eq:ComovingDetection}) we see that the condition $\Delta x^{\mu'} = 0$ can be satisfied if and only if $w^{\mu'} = v^{\mu'}$ and $\Delta \tau_c = \Delta\tau_2$.
In other words, the average tangent four-vector through the cloak must be the same as vacuum, and the time it takes for light to pass through the cloak is the same as that of an uncloaked ray, which is the condition for perfect cloaking.
Thus we see that the only way to have no image shift or distortion in both the cloak frame and the laboratory frame is if the cloak is perfect.
For all purposes the detector cannot differentiate between a wavefront passing through a perfect cloak and a wavefront passing through vacuum.

An amplitude cloak preserves the amplitude and spatial direction of throughgoing rays in the cloak frame, but introduces a phase or time delay relative to uncloaked rays traversing the same region.
In this case the average tangent four-vector is actually time-like; the light pulse travels through the cloak slower than an uncloaked pulse, and the duration of affine time required to traverse the cloak is longer than that of an uncloaked pulse, $\Delta\tau_c > \Delta\tau_2$.

Inside impedance-matched dielectric media, the tangent four-vector is no longer related to the wave four-vector by Eq.\ (\ref{Eq:VacuumTangent}).
Instead, in the rest frame of the cloak we may relate the average tangent and wave four-vectors by an effective optical metric
\begin{equation}
\rho^{\mu'\nu'} = \begin{pmatrix}
 -1 & 0 & 0 & 0 \\
 0 & (n')^{-1} & 0 & 0 \\
 0 & 0 & (n')^{-1} & 0 \\
 0 & 0 & 0 & (n')^{-1}
\end{pmatrix},
\end{equation}
where $n'$ is the amplitude cloak's effective index of refraction along the ray as determined by an observer in $S'$.
This implies that the affine time required to cross the cloak is $\Delta\tau_c = n'\Delta\tau_2$ and that
\begin{equation}
 w^{\mu'} = \left(v^{t'},\frac{v^{a'}}{n'}\right),
\end{equation} 
from which it follows that
\begin{equation}
  \Delta x^{\mu'} = w^{\mu'}\Delta\tau_c - v^{\mu'}\Delta\tau_2 = \left((n'-1) d',0,0,0\right),
\end{equation}
where $d'$ is the spatial length of the ray transect through the cloak.  More specifically, $\Delta t'_0 = v^{t'}\Delta\tau_2/c$ is the coordinate time in the cloak frame that would be required for light to transit a comparable vacuum region, and for a ray transect of spatial length $d'$, then $\Delta t_0' = d'/c$.

What this result says is that in the cloak frame the ray arrives to the same spatial coordinate point but with a time delay $\Delta t' = (n'-1)d'/c$, which agrees with our expectation for the operation of an amplitude cloak in its rest frame.
However, since the cloak is moving with respect to the laboratory frame, a late arrival to a spatial coordinate point in the cloak frame corresponds to a spatial offset in the laboratory frame in addition to a late arrival.
Indeed, boosting to the laboratory frame we find that the detection event has been Fresnel-Fizeau dragged to
\begin{equation}
 \bar{x}^{\mu} = \Lambda\indices{^{\mu}_{\mu'}} \bar{x}^{\mu'} = x^{\mu}+\Delta x^{\mu},
\end{equation}
where the displacement is
\begin{equation} \label{Eq:LorentzTransformedDisplacement}
 \Delta x^{\mu} = \Lambda\indices{^{\mu}_{\mu'}}\Delta x^{\mu'} = (n'-1)\gamma d'(1, \beta,0,0)
\end{equation}
and $\gamma = (1-\beta^2)^{-1/2}$.

What we find therefore is that in the rest frame of the amplitude cloak, a throughgoing ray ultimately reaches the same spatial coordinate point as an uncloaked ray, albeit at a later time.  
But when the cloak is in relative motion, the ray is Fresnel-Fizeau dragged to a different spatial coordinate point compared to an uncloaked ray, and the displacement depends on the amplitude cloak's effective index of refraction, its speed, and the proper spatial transit length of the ray through the cloak.
Therefore, in any frame but the rest frame of the cloak, image distortions due to Fresnel-Fizeau drag will betray the presence of the cloak.

Furthermore, it is clear from Eq.\ (\ref{Eq:LorentzTransformedDisplacement}) that for the displacement to be zero in any frame it must be zero in all frames, and that the displacement will be zero if and only if the effective index of refraction along the ray transect is $n'=1$.  
In other words that the cloak is effectively vacuum, which is only true for the single-frequency perfect cloak.

\subsection{Amplitude cloaks cannot be fine tuned for all observers}
For single-frequency perfect cloaks, an infinite number of special combinations of wave four-vector, cloak four-velocity, and cloak operational frequency satisfying Eq.\ (\ref{Eq:DispersionCondition}) exist \cite{Halimeh2016a}, but this is not possible for an amplitude cloak.
Suppose we want to fine tune a moving device such that it functions as an amplitude cloak according to the laboratory observer $S$.
By the definition of an amplitude cloak, a cloaked light pulse in $S$ should reach the detection event
\begin{equation}
\bar{x}^{\mu} = x^{\mu} + \Delta x^{\mu},
\end{equation}
where now
\begin{equation}
 \Delta x^{\mu} = ((n-1)d,0,0,0)
\end{equation}
is merely a time delay proportional to the transect length $d$ and effective index of refraction $n$ along the transect as measured by $S$.
Lorentz transforming to frame $S'$ comoving with the cloak we now find
\begin{equation}
 \bar{x}^{\mu'} = x^{\mu'} + \Delta x^{\mu'},
\end{equation}
where
\begin{equation}
 \Delta x^{\mu'} = (n-1)\gamma d(1,-\beta,0,0)
\end{equation}
now has a spatial displacement in addition to just a time delay, meaning the ray does not end up at the same spatial point as would an uncloaked ray in this frame.
Such a device, that deflects throughgoing rays to compensate for the relative motion, could surely be built but would not be interpreted as a cloak in the device rest frame because its presence would be exposed by the distortions introduced by such deflections.
Furthermore, Lorentz transforming again to any third inertial frame $S''\neq S'$ would clearly not recover the cloak functionality, and we conclude that amplitude cloaking is possible in only one choice of inertial frame and that the cloak must be tailored specifically for the choice of inertial frame and relative cloak motion.

\subsection{Colinear light rays and cloak velocity}

There does exist one exception.  When the light rays and cloak velocity are colinear it is clear that the spatial shift is in the same direction as the light propagation and therefore manifests as only a time delay on the imaging plane.

Of course it is possible to select different, more complicated imaging planes and directions of cloak motion.
Such choices will obviously change some details of the calculated shift, but the generic feature of an image shift emerges for all non-colinear combinations of light ray and cloak velocity.

\section{Ray tracing} \label{Sec:Simulations}
We now numerically illustrate the effect of Fresnel-Fizeau drag on invisibility using the invisible sphere \cite{Minano2006} as an example, while also simulating the moving free-space single-frequency perfect cylindrical cloak \cite{Pendry2006,Schurig2006a} for comparison. In this section, we shall drop the prime notation, even though we are in the cloak frame, for the sake of brevity and notational convenience. For light at its operational frequency, the cylindrical cloak exhibits an impedance-matched, anisotropic, inhomogeneous optical distribution given by 

\begin{equation}
\overset\leftrightarrow{\epsilon}=\overset\leftrightarrow{\mu}=
         \begin{pmatrix}
	\frac{r-R_1}{r} & 0 & 0 \\
	0 & \frac{r}{r-R_1} & 0 \\
	0 & 0 & \left[\frac{R_2}{R_2-R_1}\right]^2\frac{r-R_1}{r}
	\end{pmatrix},
\end{equation}

\noindent where $R_1$ and $R_2=2R_1$ are the inner and outer radii of the cylindrical cloak, respectively, and $r\in[R_1,R_2]$ is the spatial position of the wavefront within the cloak. The ray tracing of this cloak is implemented following the Hamiltonian formulation of light propagation \cite{Schurig2006a,Danner2010b} where one integrates Hamilton's equations taking $\mathcal{H}=\vec{k}^{T}\overset\leftrightarrow{\epsilon}\vec{k}-\det{\overset\leftrightarrow{\epsilon}}$ as the Hamiltonian of light.

The invisible sphere is not a cloak, but the concept is still the same, as our formalism is general for any kind of invisibility device. The refractive index distribution $n=n(r)$ of the invisible sphere \cite{Minano2006,Danner2010a,Halimeh2012a} is isotropic inhomogeneous and given implicitly by the relation

\begin{equation}
\sqrt{n}=\frac{R}{rn}+\sqrt{\left(\frac{R}{rn}\right)^2-1},
\end{equation}

\noindent with $R=R_2$ as the radius of the invisible sphere and $r\in[0,R]$ the spatial position of the wavefront within the invisible sphere. One can see that the invisible sphere is impedance-matched to vacuum with its refractive index going from unity at its circumference to infinity at its center. It adds $2\pi R$ to the optical path length of every ray that traverses it while preserving its amplitude \cite{Halimeh2012a}. Due to the infinite refractive index at its center, this device is very dispersive, and therefore has a very narrow bandwidth. However, once light is tuned in its direction of propagation and frequency such that Eq.~\eqref{Eq:DispersionCondition} is satisfied, the invisible sphere behaves similarly to an amplitude cloak for the incident light, preserving its amplitude but not its phase. For the ray tracing of the invisible sphere, we use the Newtonian formulation of light propagation in isotropic inhomogeneous media \cite{Halimeh2011} according to the equation of motion

\begin{equation}
\frac{d\vec{v}}{dt}=\frac{|\vec{v}|^2\nabla n-2(\nabla n\cdot \vec{v})\vec{v}}{n},
\end{equation}

\noindent where $\vec{v}$ is the ray velocity in the medium.

\begin{figure}[]
\includegraphics{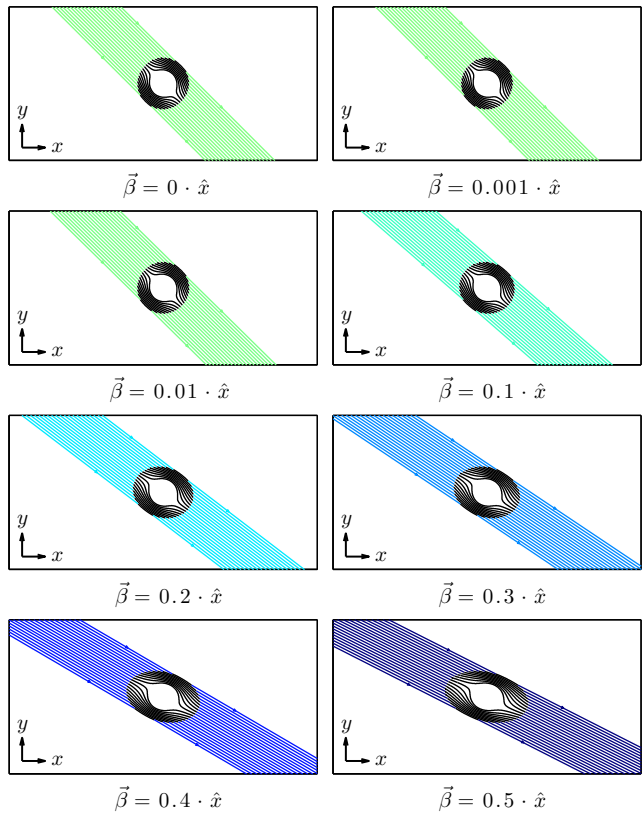}
\caption{(Color online)
A perfect cloak moving at different speeds in the positive $x$-direction of the laboratory frame is illuminated by light from the top left of each panel. All panels carry the same length scale. 
The ingoing wave four-vector $k_{\mu}$ is adjusted (blue-shifted) for each speed to ensure that Eq.\ (\ref{Eq:DispersionCondition}) is satisfied and the frequency $\omega'$ observed in the cloak frame always corresponds to the green color, which denotes the cloak operational frequency, in the top left panel, where the cloak is at rest in the laboratory frame. It is also adjusted such that the ray always enters the cloak at the same angle in the cloak rest frame.
The black parts of the light traces are its paths inside the cylindrical cloak, where the wave four-vector is not null.
Perfect cloaking is exhibited at all relative speeds.
}
\label{Fig:CC_blue}
\end{figure}

\begin{figure}[]
\includegraphics{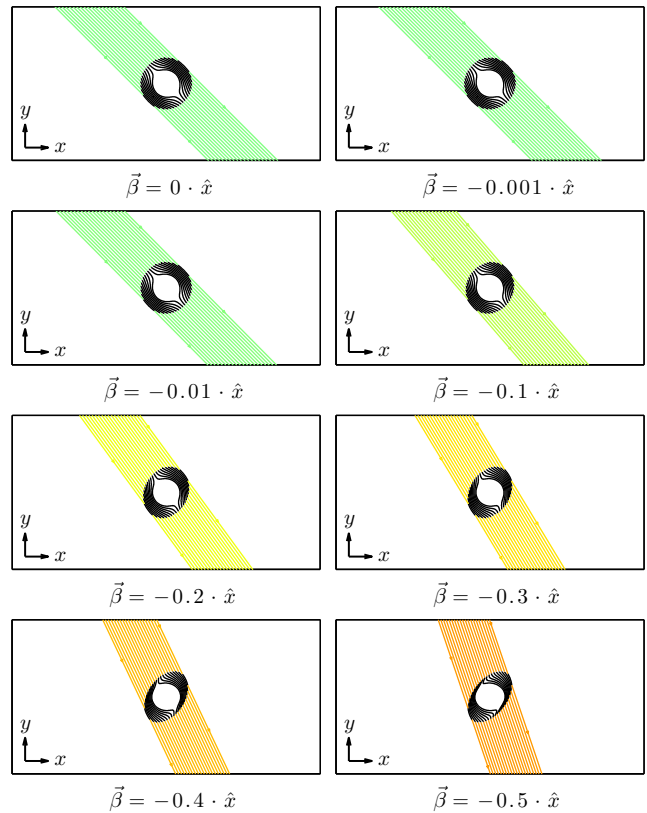}
\caption{(Color online)
Same as Fig.\ \ref{Fig:CC_blue} but for a cloak moving in the negative $x$-direction of the laboratory frame.  
The incident wave vector is adjusted (red-shifted) for each speed to ensure Eq.\ (\ref{Eq:DispersionCondition}) is satisfied and to maintain constant green color frequency $\omega'$ coincident with the operational frequency of the cylindrical cloak, and entry angle in the cloak frame.  Perfect cloaking is again exhibited at all relative speeds.
}
\label{Fig:CC_red}
\end{figure}

Ray-tracing results for a single-frequency perfect cylindrical cloak moving with various speeds $\beta$ in the laboratory frame are displayed in Figs.~\ref{Fig:CC_blue} and \ref{Fig:CC_red}.
Fig. \ref{Fig:CC_blue} depicts the cloak moving in the positive $x$-direction, while Fig.~\ref{Fig:CC_red} depicts the cloak moving in the negative $x$-direction.
For both directions of motion, light is incident from the top left of each panel and the wave four-vector $k_{\mu}$ is adjusted for each relative velocity such that a) Eq.~(\ref{Eq:DispersionCondition}) is satisfied, and b) as seen in the cloak frame, the light is always incident from the top left at $45^{\circ}$ with a frequency corresponding to the green color in the top left panel of each figure, taken to be the proper operational frequency of the cloak.  
Ray color in each panel indicates the frequency of the incident light for each speed as perceived in the laboratory frame, and is thus increasingly blue-adjusted for the receding cloak in Fig.~\ref{Fig:CC_blue}, and increasingly red-adjusted for the approaching cloak of Fig.~\ref{Fig:CC_red}.
These ray tracing simulations validate our analytical demonstration that cloaking is perfect in every inertial frame.

Since only relative velocities are meaningful, one may alternatively interpret Figs.~\ref{Fig:CC_blue} and~\ref{Fig:CC_red} as an observer either approaching or receding from, respectively, a stationary cylindrical cloak with a fixed light source illuminating it at its proper operating frequency. The approaching(receding) observer measures a blue(red)-shifted frequency, but the cloaking is still perfect at every speed. Single-frequency perfect cloaks will therefore always cloak light satisfying Eq.~\eqref{Eq:DispersionCondition} irrespective of the inertial observer.

\begin{figure}[]
\includegraphics{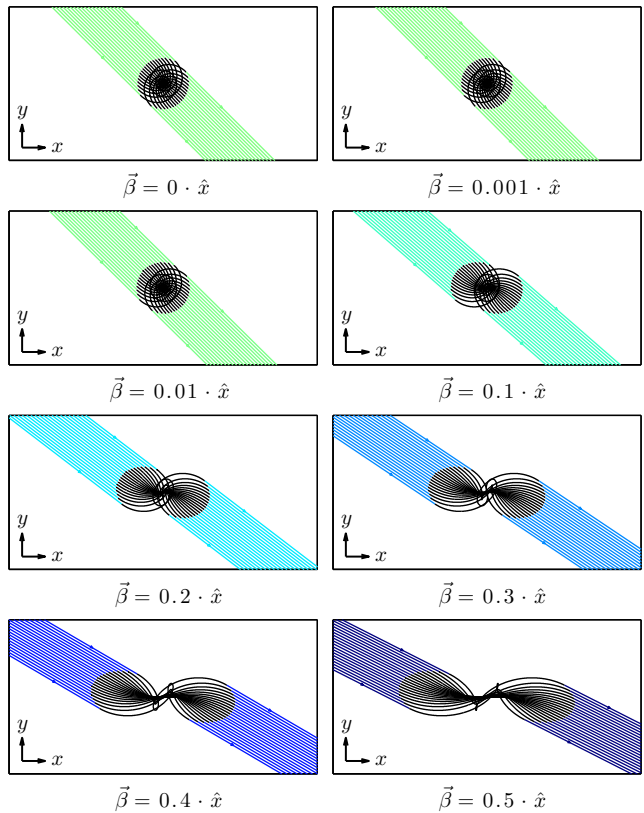}
\caption{(Color online)
As in Fig.\ \ref{Fig:CC_blue} but for an amplitude device -- an invisible sphere -- rather than a single-frequency perfect cloak.
For all non-zero speeds, light rays are displaced by Fresnel-Fizeau drag, which becomes increasingly pronounced at higher speeds.
The invisible device becomes increasingly detectable at higher relative velocities.
}
\label{Fig:IS_blue}
\end{figure}

\begin{figure}[]
\includegraphics{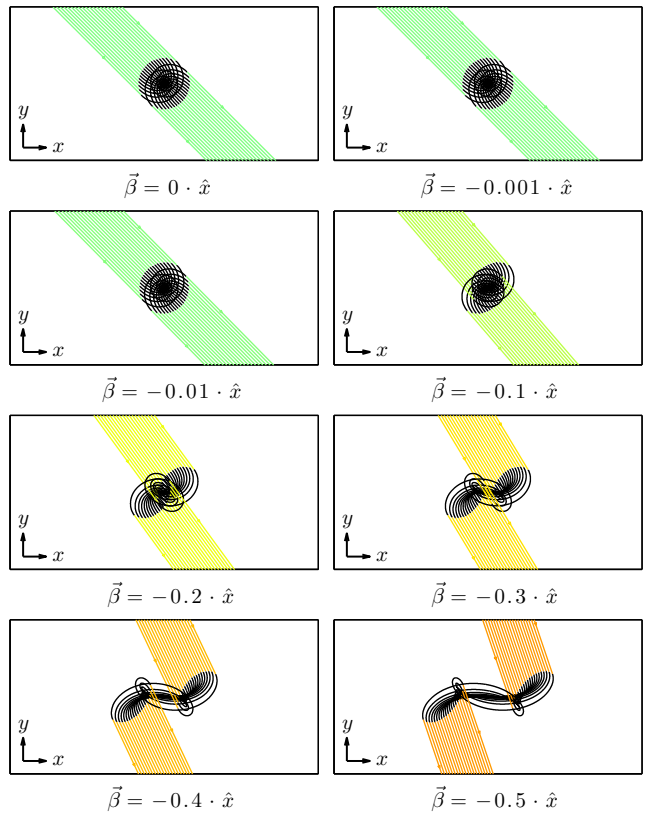}
\caption{(Color online)
As in Fig.\ \ref{Fig:IS_blue} but for an invisible sphere moving in the negative $x$-direction of the laboratory frame.  Incident light is red-shifted as in Fig.\ \ref{Fig:CC_red}.  The detailed behavior is somewhat different from Fig.\ \ref{Fig:IS_blue}, but is qualitatively the same: Fresnel-Fizeau drag distorts the image in the laboratory frame, revealing the presence of the invisible sphere.
}
\label{Fig:IS_red}
\end{figure}

Next, ray-tracing results for an invisible sphere moving with various speeds $\beta$ in the laboratory frame are displayed in Figs.~\ref{Fig:IS_blue} and \ref{Fig:IS_red}.
Fig. \ref{Fig:IS_blue} depicts the device moving in the positive $x$-direction, while Fig.~\ref{Fig:IS_red} depicts the device moving in the negative $x$-direction.
As in Figs.~\ref{Fig:CC_blue} and~\ref{Fig:CC_red}, light is incident from the top left of each panel and the wave four-vector $k_{\mu}$ is adjusted for each relative velocity such that a) Eq.~(\ref{Eq:DispersionCondition}) is satisfied, and b) as seen in the device frame, the light is always incident from the top left at $45^{\circ}$ with a frequency corresponding to the green color in the top left panel of each figure, taken to be the proper operational frequency of the device.  
Ray color in each panel indicates the frequency of the incident light for each speed as perceived in the laboratory frame, and is thus increasingly blue-adjusted for the receding cloak in Fig.~\ref{Fig:IS_blue}, and increasingly red-adjusted for the approaching cloak of Fig.~\ref{Fig:IS_red}.

As with Figs.~\ref{Fig:CC_blue} and~\ref{Fig:CC_red}, only relative motion is meaningful. So Figs.~\ref{Fig:IS_blue} and \ref{Fig:IS_red} could be alternatively interpreted as depicting an observer approaching or receding from a stationary invisible sphere illuminated by a light source incident from the top left at $45^{\circ}$ with the proper operational frequency.

These ray tracing simulations validate our analytical findings and clearly depict that in an amplitude preserving device like the invisible sphere or amplitude cloak, where phase is not preserved, Fresnel-Fizeau drag leads to image distortions whose severity increases with relative speed. 
Although the distortions become prominent at speeds $\beta\gtrsim 0.1$, any non-zero relative velocity creates image distortions that could, in principle, be detected with a sufficiently sensitive detector.
Thus, we see that even when the condition Eq.~(\ref{Eq:DispersionCondition}) is satisfied, invisibility is only possible in the inertial frame of the invisible sphere, while its presence is betrayed in all other inertial frames because it does not preserve the phase of throughgoing light.

\section{Conclusion} \label{Sec:Conclusions}

The mathematically ideal full-spectrum perfect cloak would allow for identical cloak operation and invisibility for all observers at all incident frequencies.
However, the full-spectrum perfect cloak violates causality and is therefore physically unrealizable.
Single-frequency perfect cloaks and amplitude cloaks are two causality-respecting methods to physically realize an invisibility device.
A single-frequency perfect cloak preserves both amplitude and phase of a ray by allowing superluminal ray velocity of a single frequency at the expense of heavy dispersion, while an amplitude cloak can allow finite broadband operation by preserving only the amplitude of light and not its phase, at the expense of a time delay for light propagation through the cloak relative to an uncloaked ray. 

It has been previously shown that the functionality of a single-frequency perfect cloak may be compromised by the relativistic Doppler effect, but that the cloaking effect holds for all inertial observers for an infinite number of special combinations of wave four-vector and cloak four-velocity such that the Doppler-shifted frequency coincides with the cloak's operational frequency.  However, even for these special rays the single-frequency perfect cloak is fundamentally non-reciprocal in the sense that a special ray undergoing retroreflection after passing through the cloak will no longer satisfy the necessary condition and will be scattered instead \cite{Halimeh2016a}.

For a finite-bandwidth amplitude cloak, we have now demonstrated that even if the incident wave four-vector and cloak four-velocity satisfy the condition found for single-frequency perfect cloaks, unavoidable image distortions emerge for all but a single choice of inertial observer.
These image distortions are caused by another aspect of special relativity: Fresnel-Fizeau drag by the cloak introduces a spatial offset of throughgoing rays.  It should be noted that the cause of this shift is completely unrelated to the kind of spatial offsets that result from additional structures, such as impedance-matching layers, that are sometimes imposed during the construction of cloaks \cite{Liu2009}.

As seen in Figs.\ \ref{Fig:CC_blue}, \ref{Fig:CC_red}, \ref{Fig:IS_blue}, and \ref{Fig:IS_red}, the magnitude of the image distortion grows with relative velocity and only becomes pronounced at about 10\% of the speed of light.  
But in principle the presence of an amplitude invisibility device is exposed at any relative velocity and would be detectable with sufficiently sensitive detectors.
It is fascinating that these kinds of relativistic effects can have a potentially strong impact on the functionality of cloaks and invisibility devices, and while small, it is possible that such effects may need to be accounted for in high-precision applications, even those beyond the domain of invisibility.

Finally, we can state that light cloaked in one inertial frame can be cloaked in any other inertial frame if and only if the cloaking device is a single-frequency perfect cloak as perceived by this light, which in turn means that the amplitude and phase of this light is preserved by the cloaking device in all inertial frames. As discussed, such a cloaking device would have to be single-frequency in order to not violate causality. This conclusion extends to other invisible objects and invisibility devices that are not necessarily invisibility cloaks.

\section{Acknowledgments}
The authors are grateful to Prof.\ Martin Wegener (KIT) and Prof.\ J\"{o}rg Frauendiener (U.\ Otago) for fruitful discussions and valuable comments. RTT is supported by the Royal Society of New Zealand through Marsden Fund Fast Start Grant No. UOO1219.



\begin{thebibliography}{9}

\bibitem{Leonhardt2006a}
U. Leonhardt,
\emph{Optical Conformal Mapping}.
Science \textbf{312}, 1777-1780 (2006).

\bibitem{Pendry2006}
J. B. Pendry, D. Schurig, and D. R. Smith, 
\emph{Controlling Electromagnetic Fields}.
Science \textbf{312}, 1780-1782 (2006).

\bibitem{Leonhardt2006b}
U. Leonhardt and T. G. Philbin,
\emph{General relativity in electrical engineering}. 
New J. Phys. \textbf{8}, 247 (2006).

\bibitem{Schurig2006a}
D. Schurig, J. B. Pendry, and D. R. Smith,
\emph{Calculation of material properties and ray tracing in transformation media}.
Opt. Express \textbf{14}(21), 9794-9804 (2006).

\bibitem{Sommerfeld1907}
A. Sommerfeld,
\emph{Ein Einwand gegen die Relativtheorie der Elektrodynamik und seine Beseitigung}.
Physikalische Zeitschrift \textbf{8}(23), 841-842 (1907).

\bibitem{Craeye2012}
C. Craeye and A. Bhattacharya
\emph{Rule of Thumb for Cloaking Bandwidth Based on a Wave-Packet Argument}.
IEEE Trans. Antennas Propag. \textbf{60}(7), 3516 (2012)

\bibitem{Monticone2013}
F. Monticone and A. Al\`u,
\emph{Do Cloaked Objects Really Scatter Less?}.
Phys. Rev. X \emph{3}, 041005 (2013).

\bibitem{Monticone2014}
F. Monticone and A. Al\`u,
\emph{Physical bounds on electromagnetic invisibility and the potential of superconducting cloaks}.
Phot. Nano. Fund. Appl. \textbf{12}(4), 330 (2014).

\bibitem{Schurig2006b}
D. Schurig, J. J. Mock, B. J. Justice, S. A. Cummer, J. B. Pendry, A. F. Starr, and D. R. Smith,
\emph{Metamaterial electromagnetic cloak at microwave frequencies}.
Science \textbf{314}, 977 (2006)

\bibitem{Leonhardt2009}
U. Leonhardt and T. Tyc,
\emph{Broadband invisibility by non-Euclidean cloaking}.
Science \textbf{323}, 110 (2009).

\bibitem{Perczel2011}
J. Perczel, T. Tyc, and U. Leonhardt,
\emph{Invisibility cloaking without superluminal propagation}.
New J. Phys. \textbf{13}, 083007 (2011).

\bibitem{Tyc2010}
T. Tyc, H. Chen, C. T. Chan, and U. Leonhardt,
\emph{Non-Euclidean Cloaking for Light Waves}.
IEEE J. Sel. Top. Quant. \textbf{16}, 418 (2010).

\bibitem{Xu2013}
L. Xu and H. Y. Chen,
\emph{Transformation optics with artificial Riemann sheets}.
New J. Phys. \textbf{15}, 113013 (2013).

\bibitem{Chen2013}
H. Chen, B. Zheng, L. Shen, H. Wang, X. Zhang, N. I. Zheludev, and B. Zhang,
\emph{Ray-optics cloaking devices for large objects in incoherent natural light}.
Nat. Commun. \textbf{4}, 2652 (2013).

\bibitem{Thompson2015}
R. T. Thompson, 
\emph{All Cloaks are Spacetime Cloaks}.
Proceedings of the 9th International Congress on Advanced Electromagnetic Materials in Microwaves and Optics (METAMATERIALS), pp. 502-504 (IEEE Xplore, 2015).

\bibitem{Halimeh2016a}
J. C. Halimeh, R. T. Thompson, and M. Wegener,
\emph{Invisibility cloaks in relativistic motion}.
arXiv:1510.06144, to appear in Phys. Rev. A (2016).

\bibitem{Fresnel1818}
A. J. Fresnel,
\emph{Lettre d'Augustin Fresnel \`a Francois Arago sur l'influence du mouvement terrestre dans quelques ph\'enom\`enes d'optique}.
Ann. Chim. Phys. \textbf{9}, 57 (1818).

\bibitem{Fizeau1851}
H. Fizeau,
\emph{Sur les hypoth\`eses relatives \`a l'\'ether lumineux}.
Comptes Rendus \textbf{33}, 349 (1851).

\bibitem{Jackson1999}
J. D. Jackson,
\emph{Classical Electrodynamics 3rd Ed.}.
(John Wiley, 1999).

\bibitem{Minano2006}
J. C. Mi\~nano,
\emph{Perfect imaging in a homogeneous three-dimensional region}.
Opt. Express \textbf{14}(21), 9627-9635 (2006).

\bibitem{Misner1973}
C. W. Misner, K. S. Thorne, and J. A. Wheeler,
\emph{Gravitation}.
(Macmillan, 1973).

\bibitem{Danner2010b}
A. Akbarzadeh and A. J. Danner,
\emph{Generalization of ray tracing in a linear inhomogeneous anisotropic medium: a coordinate-free approach}. 
J. Opt. Soc. Am. A \textbf{27}, 2558-2562 (2010).

\bibitem{Danner2010a}
A. J. Danner,
\emph{Visualizing invisibility: metamaterials-based optical devices in natural environments}. 
Opt. Express \textbf{18}(4), 3332-3337 (2010).

\bibitem{Halimeh2012a}
J. C. Halimeh and M. Wegener,
\emph{Time-of-flight imaging of invisibility cloaks}.
Opt. Express \textbf{20}(1), 63-74 (2012).

\bibitem{Halimeh2011}
J. C. Halimeh, R. Schmied, and M. Wegener,
\emph{Newtonian photorealistic ray tracing of grating cloaks and correlation-function-based cloaking-quality assessment}.
Opt. Express \textbf{19}(7), 6078-6092 (2011).

\bibitem{Liu2009}
R. Liu, C. Ji, J. J. Mock, J. Y. Chin, T. J. Cui, and D. R. Smith, 
\emph{Broadband Ground-Plane Cloak}.
Science \textbf{323}, 366 (2009).

\end{thebibliography}
\end{document}